\documentclass{iaus}
\usepackage{graphicx}

\newcommand{\ha}{H$\alpha${}}
\def \msun{\ifmmode{{\rm\ M}_\odot}\else{${\rm\ M}_\odot$}\fi}

\title[CC SN environments] %% give here short title %%
{The Local Environments of Core-Collapse SNe within Host Galaxies}

\author[Anderson et al.]   %% give here short author list %%
{Joseph P Anderson$^1$, Stacey M Habergham$^2$, Phil A James$^2$ \&\ M Hamuy$^1$}
\affiliation{$^1$ Departamento de Astronom\'ia, Universidad de Chile, Casilla 36-D, 
Santiago, Chile \\
$^2$Astrophysics Research Institute,
Liverpool John Moores University,
Twelve Quays House,
Egerton Wharf,
Birkenhead,
CH41 1LD\\
email: {\tt anderson@das.uchile.cl}}

\pubyear{2012}
\volume{279}  %% insert here IAU Symposium No.
\pagerange{1-4}
\jname{Deaths of Massive Stars: Supernovae and Gamma-Ray Bursts}
\editors{P. Roming, N. Kawai \&\ E. Pian}
\begin{document}

\maketitle

\begin{abstract}
We present constraints on core-collapse supernova progenitors through
observations of their environments within host galaxies. This
is achieved through 2 routes. Firstly, we investigate the
spatial correlation of supernovae with host galaxy star
formation using pixel statistics. We find that the main supernova types form
a sequence of increasing association to star formation. The most
logical interpretation is that this implies an increasing progenitor mass
sequence going from the supernova type Ia arising from the lowest mass,
through the type II, type Ib, and the supernova type Ic arising from the 
highest mass progenitors. We find the surprising result that the 
supernova type IIn show a lower association to star
formation than type IIPs, implying lower mass progenitors.
Secondly, we use host HII region spectroscopy to investigate differences
in environment metallicity between different core-collapse types.
We find that supernovae of types Ibc arise in slightly higher metallicity 
environments than type II events. However, this difference is not significant,
implying that progenitor metallicity does not play a dominant role in deciding
supernova type.

%% add here a maximum of 10 keywords, to be taken form the file <Keywords.txt>
\end{abstract}

\firstsection % if your document starts with a section,
              % remove some space above using this command.
\section{Introduction}
Mapping the links between progenitors and observed transients has
become a key goal of supernova (SN) science, helping us to understand stellar
evolution processes while also strengthening our confidence in the use of
SNe to understand the Universe. In a number of cases the
progenitor stars of core-collapse (CC) SNe have been directly identified on
pre-explosion imaging (e.g. \cite[Smartt 2009]{sma09b}). 
This allows one to gain `direct' knowledge of 
progenitor masses when observed luminosities are compared to stellar
models. However, the rarity of such detections limits the statistical results
which can be gained. One can also relate overall host galaxy properties to the 
relative rates of SNe to infer progenitor 
properties (e.g. \cite[Prieto, Stanek \&\ Beacom 2008]{pri08}, 
\cite[Boissier \&\ Prantzos 2009]{boi09}). While this can attain
significant statistics, the presence of multiple
stellar populations within galaxies 
complicates the implications which can be drawn. Our method is intermediate
to these; we investigate differences in CC SN progenitor
characteristics using observations of the exact environments at the discovery positions
of their explosions. Other recent examples using similar techniques include: \cite[Leloudas
et al. (2010)]{lel10}, \cite[Modjaz et al. (2011)]{mod11} 
and \cite[Kelly \&\ Kirshner (2011)]{kel11}.\\

\section{Spatial correlations with star formation}
\begin{figure*}
\includegraphics[width=6.8cm]{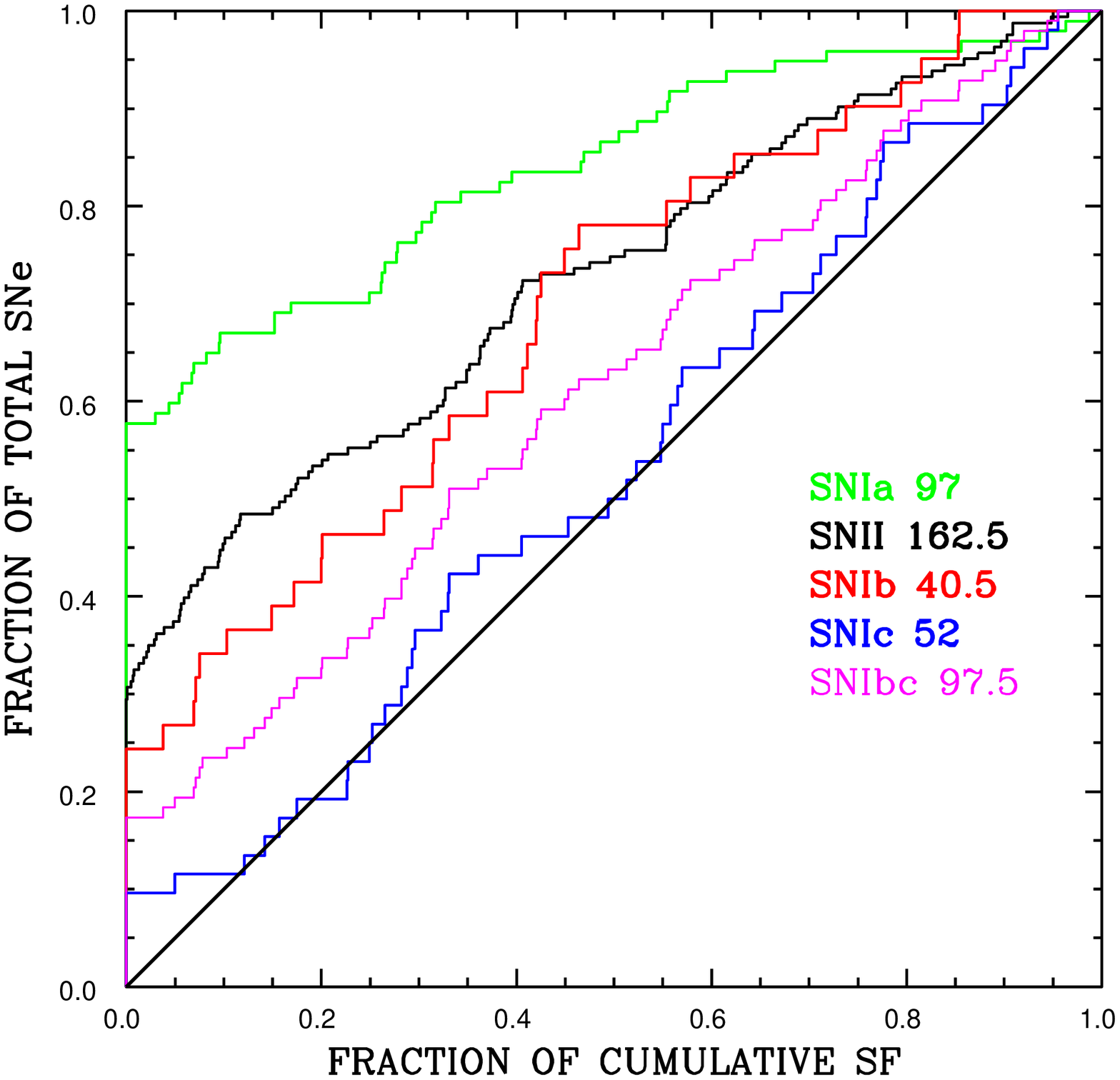}
\includegraphics[width=6.8cm]{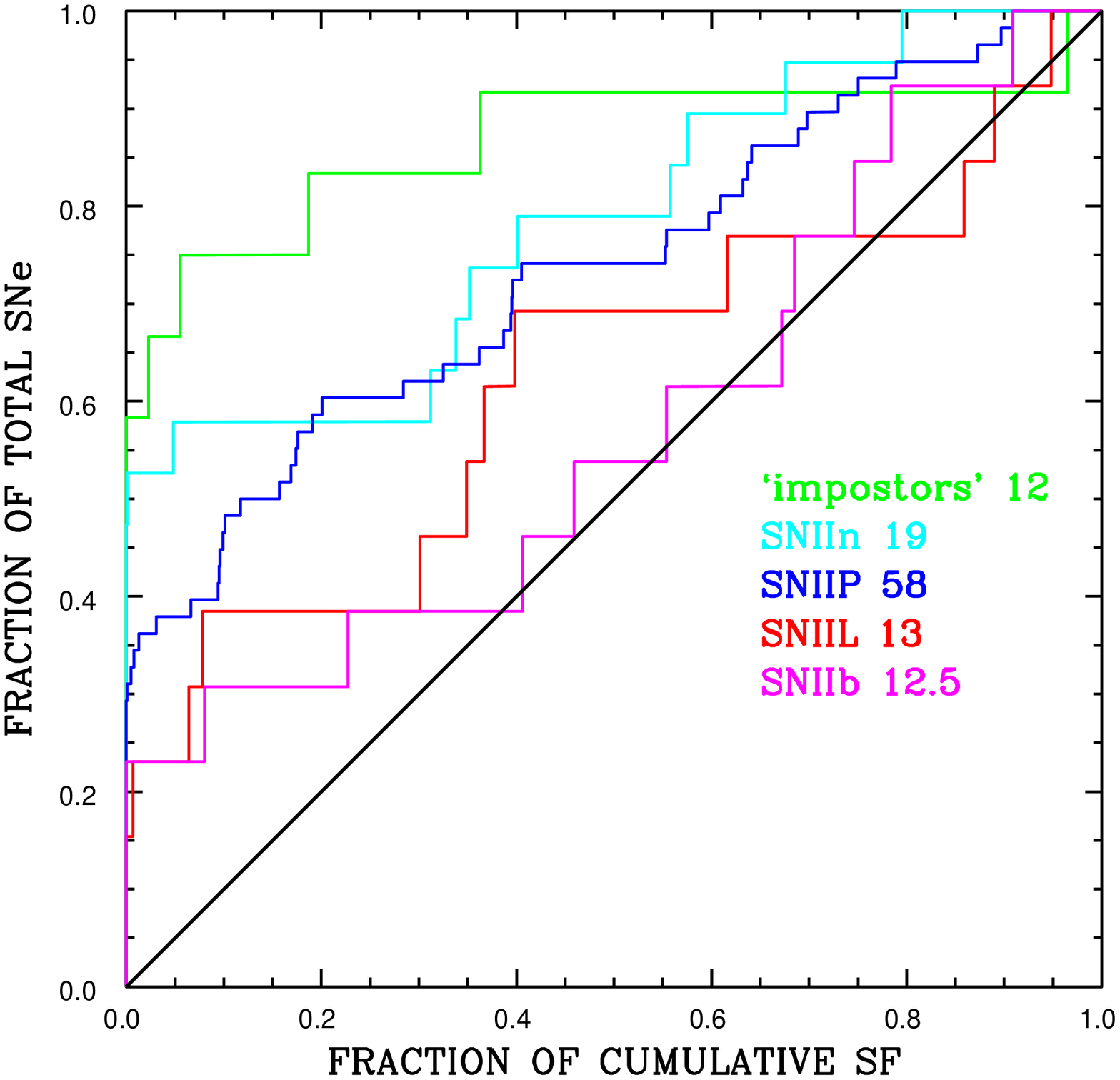}
\caption{\textbf{Left a):} cumulative pixel statistics of the main SN
types. SNIa are in green, SNII black, SNIb red and SNIc blue
(combined SNIbc magenta). The black diagonal line
illustrates a flat distribution that accurately traces the SF. 
As a distribution moves away to the left from
this diagonal it displays a lower association with SF. 
\textbf{Right b):}
Same as a) but for SNII sub-types.
`Impostors' are shown in green, SNIIn cyan, SNIIP blue, SNIIL red, and SNIIb magenta.}
\end{figure*}
We have obtained host galaxy \ha\ imaging for 260 CC SNe. The 
sample can be separated into 58 SN
type IIP (SNIIP), 13 IIL, 12.5 IIb, 19 IIn, 12 `impostors' plus 48 SNII 
(no sub-type), and 40.5 SNIb, 52 SNIc plus 5 SNIb/c.
To analyse the association of each SN with the SF of its host galaxy we use
our `NCR' pixel statistic presented in \cite[James \&\ Anderson (2006)]{jam06}. To
produce an NCR value for each SN we proceed with the following steps. 
First the sky-subtracted \ha\ image pixels are ordered into a rank of
increasing pixel count. Alongside this we form the cumulative distribution of
this ranked increasing pixel count. Negative cumulative values are then set to
zero and all other values are normalised to the total flux summed over all
pixels. It follows that each pixel has a value
between 0 and 1, where values of 0 correspond to zero flux values while
a value of 1 means the pixel has the highest count within the image. 
We measure an NCR value for each SN within our sample and proceed to
build distributions for each CC SN type.
The resulting cumulative distributions are displayed in Figure 1. In both of
these figures (a and b), as a distribution moves closer to the black diagonal
line from the top left hand side of the plots, it is showing a
higher correlation with bright HII regions. In Fig. 1a) we see a clear
sequence of increasing association to SF going from the SNIa, 
through the SNII, the SNIb and finally the
SNIc. 
The most logical way to interpret differences in associations to SF is that
more massive stars will be more highly correlated with bright HII regions. 
This is because these stars are a)
more likely to produce sufficient ionizing flux to produce bright HII regions,
and b) they will have shorter lifetimes and therefore
have less time to move away from host HII regions. The implication
is then that we observe a sequence of 
increasing progenitor mass. This starts with the SNIa
arising from the lowest mass, followed by the SNII, then the SNIb, 
and finally the SNIc arising from the most massive progenitors. 
These trends were first seen in a smaller sample published
in \cite[Anderson \&\ James (2008)]{and08}. However, the current data set is the first to
clearly separate the SNIb from the SNIc, with the latter now being seen to be
significantly more associated with SF, and hence arising from higher
mass progenitors (these results, plus discussion will be
presented in Anderson
et al. in preparation).
On the sub-types plotted in Fig. 1b) we find the interesting result that both
the `impostors' and SNIIn show a lower degree of association with SF than the
SNIIP. While there is likely to be a strong selection effect with respect to
the `impostors', for the SNIIn this implies that these events primarily arise
from relatively low mass progenitors, contrary to general consensus. Finally,
we find some suggestive constraints that both the SNIIL and SNIIb arise from
higher mass progenitors than SIIP.  

\section{Host HII region spectroscopy}
For 96 CC SNe (58 SNII and 38 SNIbc) we have also obtained host HII 
region optical spectroscopy of their immediate environments which we use to determine
region metallicities. To achieve this we
use the diagnostics of \cite[Pettini \&\ Pagel (2004)]{pet04} which give gas-phase
oxygen metallicities from emission line ratios.
In Figure 2 we show
the resulting distributions of each CC SN sample.
\begin{figure*}
\includegraphics[width=6.8cm]{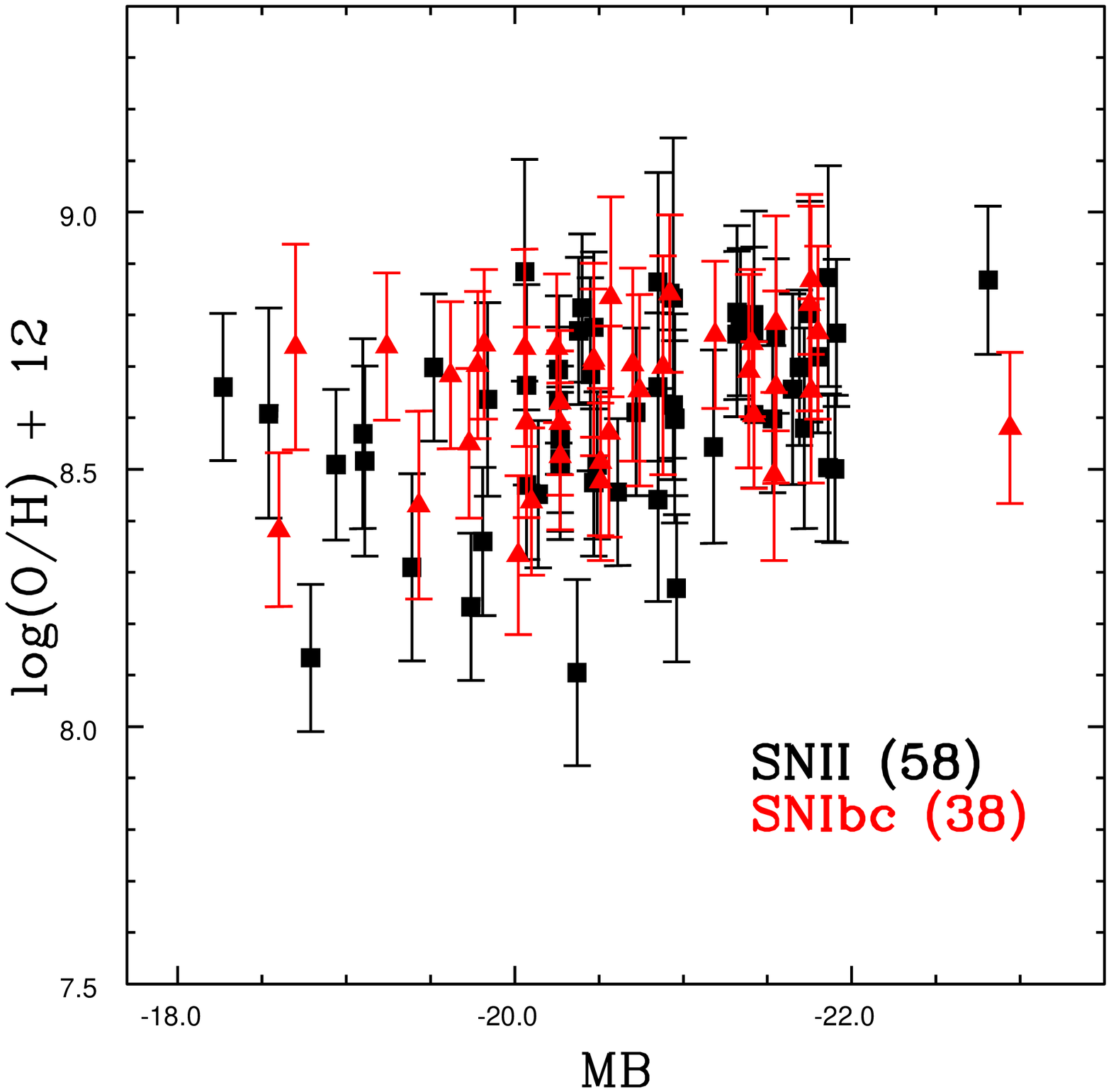}
\includegraphics[width=6.8cm]{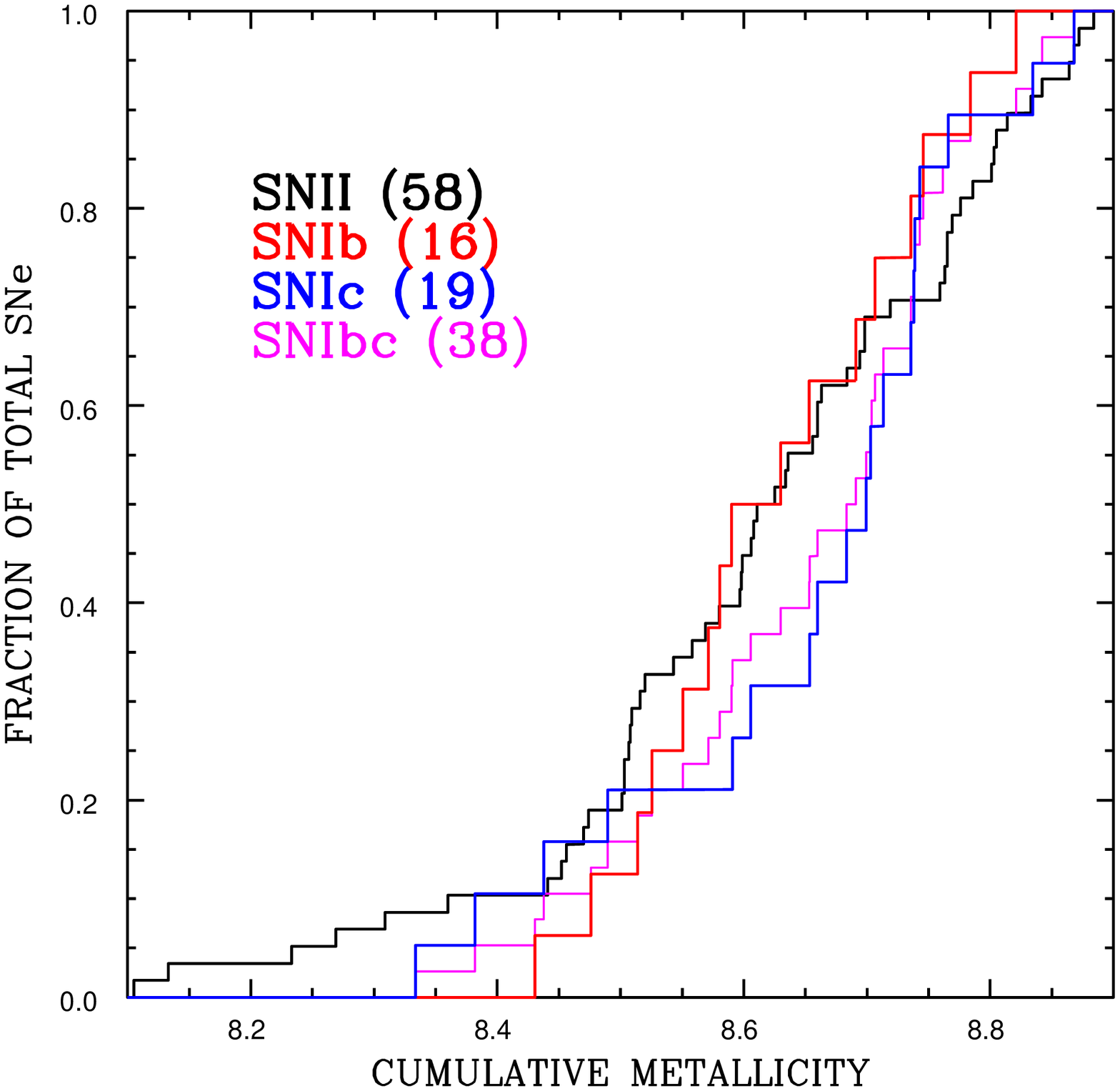}
\caption{\textbf{Left a):} Host galaxy absolute $B$-band
magnitudes against derived environment metallicities. SNII are shown in
black squares while SNIbc are shown in red triangles. \textbf{Right b):}
Cumulative metallicity distributions. The SNII distribution is
plotted in black and the SNIbc distribution is shown in
magenta. We also show the individual SNIb (red) and SNIc (blue) distributions.}
\end{figure*}
Fig. 2a) plots the absolute $B$-band magnitude of host galaxies against
derived 
environment metallicities. While the mean metallicity
of SNIbc is higher than that of the SNII, this
difference is not significant, and there is no clear offset between the
distributions. In Fig. 2b) we display the cumulative metallicity
distributions of the SNII, SNIb, SNIc and the combined SNIbc. While we
see a metallicity trend going from the SNII in the lowest, through the SNIb to the SNIc
within the highest metallicity regions, the difference between these
distributions is statistically insignificant. Overall we see that SNIbc and SNII arise
from similar metallicity environments, implying similar progenitor
metallicities. While this is the only study thus far to derive `direct'
environment metallicities to probe differences between SNII and SNIbc events
(a smaller sample was presented in \cite[Anderson et al. 2010]{and10}),
other recent results have been published on differences between SNIb, SNIc and
LGRBs, finding different results (\cite[Leloudas et al. 2011]{lel11}, 
\cite[Modjaz et al. 2011]{mod11}, and talks at this
symposium by the same authors).  

\section{Conclusions}
Through observations of the environments of CC SNe we have presented
progenitor 
constraints. We find that the main
SN types can be ordered into a sequence of increasing progenitor mass:
SNIa-SNII-SNIb-SNIc. With respect to progenitor metallicity we find no
significant difference between the SNII and SNIbc populations. This argues
that progenitor mass is the dominant (over metallicity) progenitor
characteristic that influences resulting SN type.\\

\textbf{Acknowledgments:} J.A. acknowledges support from FONDECYT grant 
3110142, and grant ICM P10-064-F (Millennium Center for
Supernova Science), with input from `Fondo de
Innovación para la Competitividad, del Ministerio de
Economía, Fomento y Turismo de Chile'.

\begin{discussion}
\discuss{Crowther}{Regarding the SNIIn results; if these are linked to LBV explosions,
locally LBVs span a wide range of mass with most shying away from HII regions
(except notably Eta Carinae!)}
\discuss{Anderson}{SNIIn show a lower association to
HII regions than SNIIP. The (to-date) progenitors of SNIIP are 8-16\msun. 
Given that our results suggest that SNIIn arise from
lower masses than SNIIP, LBV progenitors seem inconsistent with this picture.}

\discuss{Crowther}{How would your statistics differ if you limited your sample
volume, given that many SNIIP are faint and so inherently trace the lowest
mass progenitors?}
\discuss{Anderson}{We have split our sample by
host recession velocity. If there were a significant effect with
distance then we would expect NCR values to
differ with distance. We do not observe this.}

\discuss{Perley}{A significant fraction of Ics are probably dust-obscured. How
do you think dust extinction -either impacting the images (dust-lanes) or the
non-detection of some events- might affect your results?}  

\discuss{Anderson}{SNIc are strongly associated
with SF. If either \ha\ emission were being obscured,
or SNIc were missed due to dust, this would
only strengthen our result; more SNIc than we observe would be instrinsically
associated with HII regions.}

\end{discussion}


\begin{thebibliography}{}

\bibitem[Anderson \&\ James (2008)]{and08}
{Anderson, J.P. \&\ James, P.A.}, 2008
\textit{MNRAS}, 390, 1527

\bibitem[Anderson et al. (2010)]{and10}
{Anderson, J.P. et al.}, 2010
\textit{MNRAS}, 407, 2660

\bibitem[Boissier \&\ Prantzos (2009)]{boi09}
{Boissier, S. \&\ Prantzos, N.}, 2009,
\textit{A\&A}, 503, 137

\bibitem[James \&\ Anderson (2006)]{jam06}
{James, P.A. \&\ Anderson, J.P.}, 2006
\textit{A\&A}, 453, 57

\bibitem[Kelly \&\ Kirshner (2011)]{kel11}
{Kelly, P.L. \&\ Kirshner, R.P.}, 2011
\textit{arXiv}, 1110.1377

\bibitem[Leloudas et al. (2010)]{lel10}
{Leloudas, G., et al.}, 2010
\textit{A\&A}, 518, 29

\bibitem[Leloudas et al. (2011)]{lel11}
{Leloudas, G., et al.}, 2011
\textit{A\&A}, 530, 95

\bibitem[Modjaz et al. (2011)]{mod11}
{Modjaz, M., et al.} 2011
\textit{ApJ}, 731, 4

\bibitem[Pettini \&\ Pagel (2004)]{pet04}
{Pettini, M. \&\ Pagel, B.E.J.}, 2004
\textit{MNRAS}, 348, 59

\bibitem[Prieto, Stanek \&\ Beacom (2008)]{pri08}
{Prieto, J.L., Stanek, K.Z. \&\ Beacom, J.F.} 2008,
\textit{ApJ}, 673, 999

\bibitem[Smartt (2009)]{sma09b}
{Smartt, S.J.}, 2009,
\textit{ARAA}, 47, 63

\end{thebibliography}
\end{document}